\documentstyle[psfig]{l-aa}
\def\ltsima{$\; \buildrel < \over \sim \;$}
\def\lsim{\lower.5ex\hbox{\ltsima}}
\def\gtsima{$\; \buildrel > \over \sim \;$}
\def\gsim{\lower.5ex\hbox{\gtsima}}
\def\mdot {\dot M}
\newcommand{\be}{\begin{equation}}
\newcommand{\en}{\end{equation}}

\def\ergs {~erg$\,$s$^{-1}$}

\def\cmdue {~cm$^{-2}$}

\def\msole{~M_{\odot}}

\def\aa #1 #2 {A\&A, {#1}, #2}
\def\aas #1 #2 {A\&AS, {#1}, #2}
\def\araa #1 #2 {ARA\&A, {#1}, #2}
\def\mon #1 #2 {MNRAS, {#1}, #2}
\def\apj #1 #2 {ApJ, {#1}, #2}
\def\apjs #1 #2 {ApJS, {#1}, #2}
\def\apjl #1 #2 {ApJ, {#1}, #2}
\def\aj #1 #2 {AJ, {#1}, #2}
\def\nat #1 #2 {Nature, {#1}, #2}
\def\pasj #1 #2 {PASJ, {#1}, #2}
\def\pasp #1 #2 {PASP, {#1}, #2}

\title{ROSAT observations of A0538--66 during quiescence}
\author{S.~Campana\inst{1,2}}

\begin{document}

\institute{
{Osservatorio Astronomico di Brera, Via Bianchi 46, I-22055
Merate (Lc), Italy; \\ e-mail: campana@merate.mi.astro.it} 
\and
{Affiliated to I.C.R.A.}
}
\date{Received ; Accepted}

\maketitle
\label{sampout}

\begin{abstract}
We discuss serendipitous ROSAT observations of the Be transient
A0538--66, which contains the fastest known accreting X--ray pulsar
(69~ms). We show that this transient is detected several times during
quiescence at a level of about $10^{34-35}$\ergs. Considering this low
luminosity it is more likely that X--rays are produced by matter falling
onto the neutron star magnetosphere, rather than onto the star surface.
This could be the first example of an X--ray source in the ``propeller
regime".

\keywords{Stars: individual: A0538--66 -- X--ray: binaries -- accretion}
\end{abstract}

\section{Introduction}

The X--ray transient A0538--66 in the Large Magellanic Cloud contains the
accreting neutron star with the shortest known spin period (69 ms). The
periodic recurrence of most of its X--ray outbursts is highly suggestive
of a $16.7$~d (White \& Carpenter 1978) eccentric orbit ($e\gsim 0.4$)
around the Be star companion (Johnston, Griffiths \& Ward 1980; Skinner
et al. 1982). 
While in outburst, peak X--ray luminosities of $\sim 10^{39}$\ergs\ as
well as $\sim 2$~mag enhancements of the optical counterpart have been
detected (Skinner 1980). The outburst spectrum was collected by the
Monitor Proportional Counter (MPC) onboard Einstein. The best fit was
provided by a black body model with a temperature $T_{\rm bb}\sim
2.4$~keV, an emitting radius $R_{\rm bb}\sim 10^6$~cm and a column
density $N_H\sim 5\times 10^{22}$\cmdue\ (Ponman, Skinner \& Bedford
1984). The large emission radius makes unlikely that this
radiation arises as black body emission from a hot spot on the neutron
star surface; rather this component was interpreted as the emission of a
hot shell of material outside the magnetospheric radius (Ponman et al.
1984).

During the ROSAT all-sky survey two weaker outbursts from A0538--66 were
detected, with peak luminosities of $\sim 4$ and $\sim
2\times10^{37}$\ergs\ in the 0.1--2.4 keV range (Mavromatakis \& Haberl
1993). The emission was well fitted by a black body spectrum with a
softer temperature $T_{\rm bb}\sim 0.2$~keV, a characteristic radius
$R_{\rm bb}\lsim 5\times 10^7$~cm and a smaller column density
$N_H\sim 10^{20}$\cmdue. 

ASCA detected a similar outburst on Feb 3, 1995 at predicted periastron
passage at a level of $\sim 5.5 \times 10^{36}$\ergs\ (1--10 keV; Corbet
et al. 1995; Corbet 1996).  
A $3\,\sigma$ upper limit of $\sim 2\times10^{35}$\ergs\ (0.15--4.5
keV) on the quiescent luminosity of A0538--66 was determined with the
Imaging Proportional Counter onboard Einstein (Long, Helfand \&
Grabelsky 1981).

Here we discuss 25 serendipitous ROSAT Position Sensitive Proportional
Counter (PSPC) pointings containing A0538--66 in the field of view,
mainly carried out during a Large Magellanic Cloud survey. During these
observations A0538--66 was detected seven times and on two occasions the
number of photons was sufficient to extract a meaningful spectrum
(section 2).
In section 3 we discuss these observations in light of the possibility
that matter, accreting at such low rates, cannot reach the neutron star
surface but it is stopped at the magnetosphere by the centrifugal force
exerted by the corotating magnetic field (i.e. centrifugal inhibition of
accretion; see Stella et al. 1986; Campana et al. 1995). Our conclusions
are summarised in section 4. 

\section{ROSAT pointings}

As part of an on-going program aimed at studying the quiescent emission
of hard and soft X--ray transients, we select different public archive
ROSAT observations of A0538--66 obtained with the PSPC during 1992 and
1994 (see Table 1).
Most of these observations were carried out during a Large Magellanic
Cloud survey and are characterized by very short exposure times. On one
occasion a deep pointing was performed, having as a target 4U 0532--66.
All these observations were not centered at A0538--66, resulting in
off-axis angles varying between 4' and 41' (see Table 1).
 
\begin{table}
\label{log}
\caption[o]{Observations log.}
{\small
\begin{center}
\begin{tabular}{ccrrc}
Obs. ID.&    Date     & Time  & \ \ Off-axis& Orbital\\
 (ROR)  &             &(s)\ \ & \ \ (arcmin)& Phase$^*$ \\
\hline
rp400246& Jul  9, 1992& 14521& 34.2& 0.62--0.65\\
\hline
wp900538& Jul 22, 1993&  1377& 22.8& 0.36\\
wp900539& Jul 23, 1993&  1377& 25.1& 0.42\\
wp900549& Jul 24, 1993&  1328& 11.2& 0.43\\
summed  &  Jul  1993  &  4082& & 0.36--0.43\\
\hline
wp900543& Oct  7, 1993&  1093& 33.8& 0.94\\
\hline
wp900532& Nov  4, 1993&  1244& 19.2& 0.61\\
\hline
wp900530& Nov 28, 1993&   426& 23.1& 0.07\\
wp900537& Nov 28, 1993&   363& 13.4& 0.07\\
wp900531& Nov 29, 1993&   536& 26.7& 0.13\\
wp900544& Nov 29, 1993&  2121& 30.1& 0.13--0.17\\
wp900545& Nov 29, 1993&   363& 31.9& 0.17\\
wp900534& Nov 30, 1993&   295&  8.7& 0.17\\
wp900551& Dec  1, 1993&   990& 18.6& 0.29\\
wp900540& Dec  2, 1993&  1005& 37.8& 0.35\\
wp900552& Dec  3, 1993&  1071&  5.3& 0.41\\
wp900553& Dec  3, 1993&  1071& 26.4& \ \ \ \ \ $^{\dag}$\\
wp900533& Dec  6, 1993&  1579& 25.1& 0.57\\
summed  & Nov-Dec 1993&  8749& & 0.07--0.57\\
\hline
wp900541& Jun  5, 1994&  1002& 39.1& 0.46\\
wp900542& Jun  5, 1994&  1000& 41.4& 0.46\\
 summed &   Jun 1994  &  2002& & 0.46 \\
\hline
wp900550& Jul  4, 1994&   970& 15.9& 0.21\\
wp900535& Jul  5, 1994&   855& 13.5& 0.24\\
wp900536& Jul  6, 1994&  1006&  8.4& 0.30\\
wp900547& Jul  6, 1994&   774& 18.6& 0.30\\
wp900546& Jul  6, 1994&   596& 15.5& 0.31\\
wp900548& Jul  6, 1994&  1071&  4.1& 0.31\\
 summed &   Jul 1994  &  5272& & 0.21--0.31\\
\hline
\end{tabular}
\end{center}  }
\smallskip
\noindent $^*$ Orbital phases were derived by the ephemerides of Skinner 
(1981; see text). The error on the orbital phase is about $\pm 0.1$.

\noindent $^{\dag}$ This observation consists of two pointings separated by
$\sim 20$ days and was not considered in the present analysis.
\end{table}

The data were analysed with the XANADU package: first the event files
were taken from the ROSAT archive at the Max Planck Institute
(M\"unchen) and images were extracted, centered on the transient source
position (Johnston et al. 1980). 
For each exposure the orbital phase was derived by using the X--ray outbursts
ephemeridis by Skinner (1981): the orbital period is 
$P_{\rm orb}=16.6515\pm0.0005$ d and zero phase occurs at MJD 2443423.96
$\pm$ 0.05 (see Table 1). These uncertainties result in a typical
error on the orbital phase of about $\pm0.1$.

The exposure maps were linearly interpolated and rescaled so to
be overlaid on the images. Exposure corrected count rates or $3\,\sigma$
upper limits were derived. Most of the observations have exposure
times shorter than 1000 s, so that different images, when close in time,
were summed together in order to increase the signal to noise: 8 exposure
corrected images were the final product of this procedure.
A sliding-box detection algorithm was then applied to these summed images
revealing A0538--66 seven times with a Poisson probability to be a
background fluctuation smaller than $10^{-6}$ (see Table 2). 

The highest count rates during a single observation were achieved on Jul
1992 ($\sim 0.03$ c s$^{-1}$) and on Oct 1993 ($\sim 0.4$ c
s$^{-1}$). For these two observations the number of photons collected was
sufficient to extract a spectrum.

The spectrum from the ROR 900543 (Oct 7, 1993) observation has been
extracted from a circle of $\sim 4'$ centered on the source and the
background from a concentric annulus with inner and outer radii of 13'
and 22', respectively. 
The 303 photons were rebinned into 11 energy channels in order to have at
least 25 photons in each channel. 

In Table 3 we report the results of the fitting for different models as
well as the 90\% confidence intervals. All single component spectral
models provide a relative good fit to the data: a black body model with
$T_{\rm bb}=0.2$ keV, $R_{\rm bb}=7\times10^6$ cm and $N_H= 3\times
10^{20}$\cmdue\ a reduced $\chi^2=0.8$ (see Fig. 1); the 0.1--2.4 keV
flux amounts to $\sim 5\times 10^{-12}$\ergs\cmdue, and, assuming a
distance of 50 kpc, the luminosity to $\sim 10^{36}$\ergs. 
A bremsstrahlung model with a temperature of $T_{\rm br}=0.8$ keV and
$N_H=7\times 10^{20}$\cmdue\ gives reduced $\chi^2=0.9$; the 0.1--2.4 keV
luminosity is slightly higher $\sim 3\times 10^{36}$\ergs. 
In the case of a black body spectrum virtually all the observed flux falls inside
the ROSAT PSPC energy range; for a bremsstrahlung spectrum the bolometric correction 
is about 30\% for an energy range of 0.001 -- 100 keV. 
A bolometric correction of about an order of magnitude is expected for a power-law 
model in the same energy range. If this is the case, a large fraction of the total 
luminosity would be emitted in the optical-UV band overwhelming the 
Be companion, so that the low energy cut off should be at lower energies and
the bolometric correction smaller. 
Moreover, the power-law model provides a slightly worse fit to the ROSAT spectrum. 

%
\message{S-Tables Macro v1.0, ACS, TAMU (RANHELP@VENUS.TAMU.EDU)}
%
%
\newhelp\stablestylehelp{You must choose a style between 0 and 3.}%
\newhelp\stablelinehelp{You should not use special hrules when stretching
a table.}%
\newhelp\stablesmultiplehelp{You have tried to place an S-Table inside another
S-Table.  I would recommend not going on.}%
%
%
\newdimen\stablesthinline
\stablesthinline=0.4pt
\newdimen\stablesthickline
\stablesthickline=1pt
%
%
\newif\ifstablesborderthin
\stablesborderthinfalse
\newif\ifstablesinternalthin
\stablesinternalthintrue
\newif\ifstablesomit
\newif\ifstablemode
\newif\ifstablesright
\stablesrightfalse
%
%
\newdimen\stablesbaselineskip
\newdimen\stableslineskip
\newdimen\stableslineskiplimit
%
%
\newcount\stablesmode
\newcount\stableslines
\newcount\stablestemp
\stablestemp=3
\newcount\stablescount
\stablescount=0
\newcount\stableslinet
\stableslinet=0
%
%
%
\newcount\stablestyle
\stablestyle=0
%
%
\def\stablesleft{\quad\hfil}%
\def\stablesright{\hfil\quad}%
%
%
\catcode`\|=\active%
%
%
\newcount\stablestrutsize
\newbox\stablestrutbox
\setbox\stablestrutbox=\hbox{\vrule height10pt depth5pt width0pt}
\def\stablestrut{\relax\ifmmode%
                         \copy\stablestrutbox%
                       \else%
                         \unhcopy\stablestrutbox%
                       \fi}%
%
%
\newdimen\stablesborderwidth
\newdimen\stablesinternalwidth
\newdimen\stablesdummy
\newcount\stablesdummyc
\newif\ifstablesin
\stablesinfalse
%
%
\def\begintable{\stablestart%
  \stablemodetrue%
  \stablesadj%
  \halign%
  \stablesdef}%
\def\begintableto#1{\stablestart%
  \stablemodefalse%
  \stablesadj%
  \halign to #1%
  \stablesdef}%
\def\begintablesp#1{\stablestart%
  \stablemodefalse%
  \stablesadj%
  \halign spread #1%
  \stablesdef}%
\def\stablesadj{%
  \ifcase\stablestyle%
    \hbox to \hsize\bgroup\hss\vbox\bgroup%
  \or%
    \hbox to \hsize\bgroup\vbox\bgroup%
  \or%
    \hbox to \hsize\bgroup\hss\vbox\bgroup%
  \or%
    \hbox\bgroup\vbox\bgroup%
  \else%
    \errhelp=\stablestylehelp%
    \errmessage{Invalid style selected, using default}%
    \hbox to \hsize\bgroup\hss\vbox\bgroup%
  \fi}%
\def\stablesend{\egroup%
  \ifcase\stablestyle%
    \hss\egroup%
  \or%
    \hss\egroup%
  \or%
    \egroup%
  \or%
    \egroup%
  \else%
    \hss\egroup%
  \fi}%
\def\stablestart{%
  \ifstablesin%
    \errhelp=\stablesmultiplehelp%
    \errmessage{An S-Table cannot be placed within an S-Table!}%
  \fi
  \global\stablesintrue%
  \global\advance\stablescount by 1%
  \message{<S-Tables Generating Table \number\stablescount}%
  \begingroup%
  \stablestrutsize=\ht\stablestrutbox%
  \advance\stablestrutsize by \dp\stablestrutbox%
  \ifstablesborderthin%
    \stablesborderwidth=\stablesthinline%
  \else%
    \stablesborderwidth=\stablesthickline%
  \fi%
  \ifstablesinternalthin%
    \stablesinternalwidth=\stablesthinline%
  \else%
    \stablesinternalwidth=\stablesthickline%
  \fi%
  \tabskip=0pt%
  \stablesbaselineskip=\baselineskip%
  \stableslineskip=\lineskip%
  \stableslineskiplimit=\lineskiplimit%
  \offinterlineskip%
  \def\borderrule{\vrule width \stablesborderwidth}%
  \def\internalrule{\vrule width \stablesinternalwidth}%
  \def\thinline{\noalign{\hrule height \stablesthinline}}%
  \def\thickline{\noalign{\hrule height \stablesthickline}}%
  \def\trule{\omit\leaders\hrule height \stablesthinline\hfill}%
  \def\ttrule{\omit\leaders\hrule height \stablesthickline\hfill}%
  \def\tttrule##1{\omit\leaders\hrule height ##1\hfill}%
  \def\stablesel{&\omit\global\stablesmode=0%
    \global\advance\stableslines by 1\borderrule\hfil\cr}%
  \def\el{\stablesel&}%
  \def\elt{\stablesel\thinline&}%
  \def\eltt{\stablesel\thickline&}%
  \def\elttt##1{\stablesel\noalign{\hrule height ##1}&}%
  \def\elspec{&\omit\hfil\borderrule\cr\omit\borderrule&%
              \ifstablemode%
              \else%
                \errhelp=\stablelinehelp%
                \errmessage{Special ruling will not display properly}%
              \fi}%
  \def\stmultispan##1{\mscount=##1 \loop\ifnum\mscount>3 \stspan\repeat}%
  \def\stspan{\span\omit \advance\mscount by -1}%
  \def\multicolumn##1{\omit\multiply\stablestemp by ##1%
     \stmultispan{\stablestemp}%
     \advance\stablesmode by ##1%
     \advance\stablesmode by -1%
     \stablestemp=3}%
  \def\multirow##1{\stablesdummyc=##1\parindent=0pt\setbox0\hbox\bgroup%
    \aftergroup\emultirow\let\temp=}
  \def\emultirow{\setbox1\vbox to\stablesdummyc\stablestrutsize%
    {\hsize\wd0\vfil\box0\vfil}%
    \ht1=\ht\stablestrutbox%
    \dp1=\dp\stablestrutbox%
    \box1}%
  \def\stpar##1{\vtop\bgroup\hsize ##1%
     \baselineskip=\stablesbaselineskip%
     \lineskip=\stableslineskip%
     \lineskiplimit=\stableslineskiplimit\bgroup\aftergroup\estpar\let\temp=}%
  \def\estpar{\vskip 6pt\egroup}%
  \def\stparrow##1##2{\stablesdummy=##2%
     \setbox0=\vtop to ##1\stablestrutsize\bgroup%
     \hsize\stablesdummy%
     \baselineskip=\stablesbaselineskip%
     \lineskip=\stableslineskip%
     \lineskiplimit=\stableslineskiplimit%
     \bgroup\vfil\aftergroup\estparrow%
     \let\temp=}%
  \def\estparrow{\vfil\egroup%
     \ht0=\ht\stablestrutbox%
     \dp0=\dp\stablestrutbox%
     \wd0=\stablesdummy%
     \box0}%
  \def|{\global\advance\stablesmode by 1&&&}%
  \def\|{\global\advance\stablesmode by 1&\omit\vrule width 0pt%
         \hfil&&}%
  \def\vt{\global\advance\stablesmode by 1&\omit\vrule width \stablesthinline%
          \hfil&&}%
  \def\vtt{\global\advance\stablesmode by 1&\omit\vrule width \stablesthickline%
          \hfil&&}%
  \def\vttt##1{\global\advance\stablesmode by 1&\omit\vrule width ##1%
          \hfil&&}%
  \def\vtr{\global\advance\stablesmode by 1&\omit\hfil\vrule width%
           \stablesthinline&&}%
  \def\vttr{\global\advance\stablesmode by 1&\omit\hfil\vrule width%
            \stablesthickline&&}%
  \def\vtttr##1{\global\advance\stablesmode by 1&\omit\hfil\vrule width ##1&&}%
  \stableslines=0%
  \stablesomitfalse}
\def\stablesdef{\bgroup\stablestrut\borderrule##\tabskip=0pt plus 1fil%
  &\stablesleft##\stablesright%
  &##\ifstablesright\hfill\fi\internalrule\ifstablesright\else\hfill\fi%
  \tabskip 0pt&&##\hfil\tabskip=0pt plus 1fil%
  &\stablesleft##\stablesright%
  &##\ifstablesright\hfill\fi\internalrule\ifstablesright\else\hfill\fi%
  \tabskip=0pt\cr%
  \ifstablesborderthin%
    \thinline%
  \else%
    \thickline%
  \fi&%
}%
\def\endtable{\advance\stableslines by 1\advance\stablesmode by 1%
   \message{- Rows: \number\stableslines, Columns:  \number\stablesmode>}%
   \stablesel%
   \ifstablesborderthin%
     \thinline%
   \else%
     \thickline%
   \fi%
   \egroup\stablesend%
\endgroup%
\global\stablesinfalse}
%
%

\begin{table*}
\label{summ}
\stablesthinline=0pt
\stablesborderthintrue
\stablestyle=0
\caption{Summary of detections and upper limits for A0538--66.}
\begintable
   Date         \|  Time \| Orbital  \|          Count rate            \| Poisson         \| Luminosity      \el
                \|  (s)  \|  Phase   \|      (c s$^{-1}$)              \| probability     \| 0.1--2.4 keV (erg s$^{-1}$)\elt
Jul  9, 1992    \|  14521\|0.62--0.65\|$(3.3\pm 0.3)\,\times 10^{-2}$  \|$< 10^{-15}$     \|$1\times10^{35}$ \el
Jul  1993$^*$   \|\  4082\|0.36--0.43\| $(9.1\pm 2.2)\,\times 10^{-3}$ \|$2\times10^{-8}$ \|$3\times10^{34}$ \el
Oct  7, 1993    \|\  1093\|   0.94   \| $(3.9\pm 0.2)\,\times 10^{-1}$ \|$< 10^{-15}$     \|$1\times10^{36}$ \el
Nov  4, 1993    \|\  1244\|   0.61   \|$(1.6\pm 0.4)\,\times 10^{-2}$  \|$4\times10^{-7}$ \|$6\times10^{34}$ \el
Nov-Dec 1993$^{*,\,\dag}$\|\  8749\|0.07--0.57\|$(2.0\pm 0.5)\,\times 10^{-2}$\|$4\times10^{-13}$\|$7\times10^{34}$\el
  Jun 1994$^*$  \|\  2002\|   0.46   \| $  \lsim\,3.5\times 10^{-2}$   \|$1\times10^{-3}$ \|$\lsim1\times10^{35}$ \el
  Jul 1994$^*$  \|\  5272\|0.21--0.31\| $(1.1\pm 0.2)\,\times 10^{-2}$ \|$< 10^{-15}$     \|$4\times10^{34}$ \el
\endtable
\smallskip
\noindent The upper limit to the count rate is at a level of
$3\,\sigma$. The luminosities were derived assuming a count rate to flux
conversion of 1 c s$^{-1}=1.2\times10^{-11}$\ergs\cmdue\ based on
spectral fits (black body model; in the case of a bremsstrahlung model 
the luminosities increase by a factor of $\sim 2$).\\
\noindent $^*$ Summed images.\\
\noindent $^{\dag}$ The major contributions to the detection do not 
come from the pointings near periastron.
\end{table*}

The source spectrum of the Jul 9, 1992 observation (ROR 400246) has been
extracted from a circle of $\sim 3'$ centered at the source position; the
background has been estimated from a concentric annular region with inner
and outer radius of 13' and 22', respectively.
A smaller extraction radius than for ROR 900543 has been adopted in this
case because the source was weaker and the background stronger.
The 333 photons were rebinned into 12 energy channels in order to have at
least 25 photons in each channel. 

The spectrum is rather poor (see Table 3) and can be fitted by a black
body with $T_{\rm bb}=0.3$ keV, $R_{\rm bb}\lsim 7\times10^6$ cm and
$N_H=10^{21}$\cmdue\ (reduced $\chi^2=1.0$); the 0.1--2.4 keV flux
amounts to $\sim 4\times 10^{-13}$\ergs\cmdue\ and the luminosity to
$\sim 10^{35}$\ergs. A bremsstrahlung spectrum with $T_{\rm br}=1.0$ keV 
and a $N_H\sim 3\times 10^{21}$\cmdue\ (reduced $\chi^2=1.0$); the 
0.1--2.4 keV luminosity is $\sim 3\times 10^{35}$\ergs. The same bolometric 
corrections as before should be applied.

\begin{table*}
\label{spe}
\stablesthinline=0pt
\stablesborderthintrue
\stablestyle=0
\caption{Summary of spectral fits for A0538--66.}
\begintable
\multicolumn4 \hfill ROR 900543 (Oct 7, 1993) \hfill \elt
Model          | Column density      | Parameter                           | Red. $\chi^2$\el
               |($10^{20}$ cm$^{-2}$)|                                     |              \elt
Black body$^{\dag}$|$3.3^{+4.9}_{-2.1}$|$T_{\rm bb}=0.22^{+0.07}_{-0.04}$ keV| 0.8      \el
Bremsstrahlung |$7.5^{+7.3}_{-2.7}$  |$T_{\rm br}=0.81^{+2.01}_{-0.37}$ keV| 0.9        \el
Power law      |$9.4^{+12.3}_{-4.0}$ |$\alpha=2.5^{+1.2}_{-0.9}$           | 1.0        \el
Raymond-Smith  |$1.9^{+1.5}_{-1.8}$  |$T_{\rm RS}=0.85^{+0.25}_{-0.15}$ keV| 2.7        \el
Disk Black body|$5.9^{+5.9}_{-2.3}$  |$T_{\rm dd}=0.33^{+0.21}_{-0.09}$ keV| 0.9        \eltt
\multicolumn4 \hfill ROR 400246 (Jul 9, 1992) \hfill \elt
Model          | Column density      | Parameter                           | Red. $\chi^2$\el
               |($10^{21}$ cm$^{-2}$)|                                     |            \elt
Black body$^{\dag}$|$1.3^{+9.9}_{-0.9}$|$T_{\rm bb}=0.31^{+0.34}$ keV      | 1.0    \el
Bremsstrahlung |$2.8_{-0.7}$         |$T_{\rm br}=0.98_{-0.82}$ keV        | 1.0        \el
Power law      |$4.0$                |$\alpha=3.0_{-2.9}$                  | 1.0        \el
Raymond-Smith  |$15.1^{+4.5}_{-7.1}$ |$T_{\rm RS}=0.23^{+0.03}_{-0.23}$ keV| 1.4        \el
Disk Black body|$2.3^{+10.7}_{-0.7}$ |$T_{\rm dd}=0.42^{+2.06}_{-0.42}$ keV| 1.0        \endtable

\smallskip
\noindent $^{\dag}$ The black body radius for the Oct 1993 observation is
$R_{\rm bb}=6.9^{+3.0}_{-2.1}\times 10^6$ cm; for the Jul 1992 the black
body radius has large uncertainties, we derive $R_{\rm bb}\lsim 
7.3\times 10^6$ cm.
\end{table*}

A careful analysis at the neutron star spin period was carried out. The
upper limit on the X--ray luminosity derived by Long et al. (1981)
informs us that if matter acts to spin up the neutron star (and no
outbursts occurred) the spin period will not change significantly. On the
contrary, if during quiescence the system is in the propeller regime (see
below) large spin downs are predicted: by taking the strongest version of
the propeller regime (cf. Wang \& Robertson 1981), we derive that the
spin period of A0538--66 should not exceed 75 ms.  We take the same PSPC
counts used for the spectral analysis, after correcting their arrival
times to the solar system barycenter. No significant periodicities were
found in the range 68--75 ms providing a $3\,\sigma$ upper limit of $\sim
85\%$ on the amplitude of a sinusoidal X--ray modulation. However, the pulsed 
component amplitude of A0538--66 detected by the Einstein MPC was of $\sim 25\%$
(Skinner 1982).

\begin{figure*}
\label{br}
\centerline{\psfig{figure=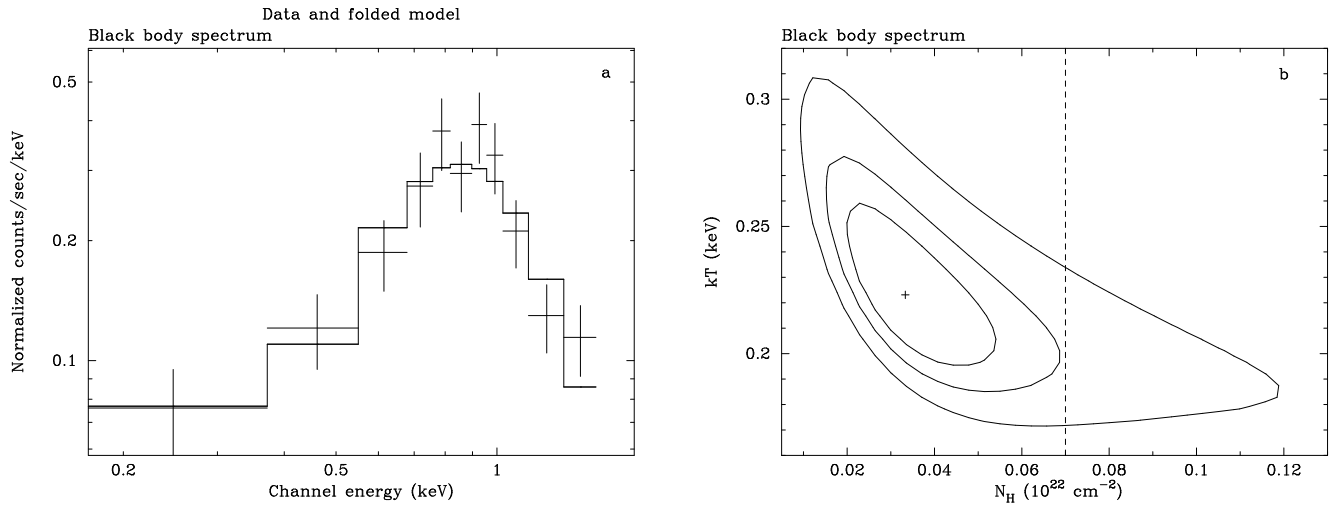,height=10.cm}}
\caption{In panel $a$ we report the spectrum of A0538--66 during the Oct,
7 1993 observation. The best fit model is provided by a black body
spectrum. In panel $b$ the confidence contours (1, 2 and 3 $\sigma$) for
the black body model are reported. The cross signs the best fit model. The
dashed line marks the mean galactic column density towards the Large
Magellanic Cloud.}
\end{figure*}

\section{Accretion in the propeller regime}

A0538--66 is a transient X--ray source, the outburst of which likely derive
from an enhanced emission of its Be companion star (hard X--ray
transients). In the previous section we presented the first detection
of A0538--66 during quiescence. The detected luminosity could in
principle derive from different mechanisms (cf. Stella et al. 1994),
its emission level is however sufficiently high not to be accounted by
the Be companion star emission (which should emit $\sim 10^{32}$\ergs\
at most; Meurs et al. 1992), nor by the emission of the underlying
neutron star either due to cooling or by non-thermal processes. The
most likely explanation is that the observed luminosity derives from
accretion. Two different regimes are possible in this case (e.g.
Illarionov \& Sunyaev 1975). The motion of the matter falling in the
potential well of a neutron star becomes dominated by the rapidly
increasing magnetic pressure at the magnetospheric radius, $r_{\rm m}$.
At smaller radii the accretion flow follows the magnetic field lines and is
enforced to corotate with the neutron star. If the magnetospheric
radius lies inside the corotation radius ($r_{\rm cor}$), i.e. the
radius at which a test particle in Keplerian orbit would corotate with
the neutron star, the accreting matter can proceed down to the neutron
star surface. 

Due to the scaling of $r_{\rm m}$ with the accretion rate a minimum
value, $\mdot_{\rm min}$, exists below which centrifugal acceleration is
strong enough to expels the infalling matter, preventing the accretion.
This limiting accretion rate $\mdot_{\rm min}$ corresponds to a minimum
accretion luminosity down to the neutron star surface of
\begin{eqnarray} 
L_{\rm min} & = & {{G\,M\,\mdot_{\rm min}}\over {R}} \nonumber \\ 
& = & 8\times 10^{38}\, \mu_{29}^2\,M_{1.4}^{-2/3}\,P_{69\, \rm
ms}^{-7/3}\,R_6^{-1} ~{\rm erg\,s^{-1}} \ , 
\label{lmin} 
\end{eqnarray}
\noindent where $P_{69\,{\rm ms}}$ is the neutron star spin period in
units of 69 ms and $M_{1.4}$ its mass in units of $1.4\,\msole$. 
The neutron star magnetic moment $\mu=B\,R^3/2$ is in units of
$\mu=10^{29}\,\mu_{29}$ G$\,$cm$^{3}$, being $R=10^6\,R_6$ cm the neutron
star radius and $B$ the surface magnetic field (e.g. Stella et al. 1994). 

The high X--ray luminosity ($\sim 10^{39}$\ergs; Skinner et al. 1980) and
the presence of pulsations testify that during the bright outbursts the
centrifugal barrier is open and accretion onto the neutron star surface
takes place. As noted by Skinner et al. (1982), the lower range of X--ray
luminosities observed during bright outbursts implies an upper limit of
$\mu_{29} \lsim 1$.
On the other hand, the presence of X--ray pulsations indicates that 
a small magnetosphere is still present (i.e. $r_{\rm m} \!>\! R$); this 
implies $\mu_{29} \gsim 3\times 10^{-3}$. 

Below the minimum luminosity $L_{\rm min}$, the accretion flow is halted
at the magnetospheric boundary, which behaves like a closed barrier.
Entering the propeller regime a drop in the accretion-induced luminosity
is expected to occur down to 
\be
L_{\rm cor}={{G\,M\mdot_{\rm min} }\over {r_{\rm cor}}} \sim 3\times
10^{37}\,\mu_{29}^{2}\,M_{1.4}^{-1}\,
P_{69\,{\rm ms}}^{-3} {\rm \ erg\,s^{-1}} \label{lcor}
\en
\noindent which is the maximum allowed luminosity in this regime; for 
lower accretion rates the emitted luminosity scales as 
$(\mdot/\mdot_{\rm min})^{9/7}$ (for more details see Campana et al. 
1995; Corbet 1996).

If the minimum detected X--ray luminosity ($\sim 5\times10^{34}$\ergs,
for a black body model) derives from accretion onto the neutron star surface, 
we have from eq. 1 an upper limit to the magnetic moment of $\mu_{29}\lsim
8\times10^{-3}$. This field is likely too low to explain the observed
pulsations during outburst. On the other hand, no stringent limitations on
the magnetic moment exist if the accretion matter is stopped at the
magnetospheric radius by the centrifugal barrier (one has to require
$\mu_{29} \lsim 10$ so that a radio pulsar cannot turn on; see
Campana et al. 1995).

\section{Conclusions} 

In Campana et al. (1995) we suggested that the low-luminosity ($\lsim
10^{38}$\ergs) and soft outbursts of A0538--66 (cf. Mavromatakis \&
Haberl 1993) could be powered by accretion onto the neutron star
magnetosphere, whereas during the outbursts with harder spectra and
$L\gsim 10^{38}$\ergs\ (when also the 69~ms pulsations were detected) the
centrifugal barrier is open and the accretion flow can proceed down to
the neutron star surface. This interpretation implies $\mu_{29}\sim 1$. 

In this paper we report several detections of A0538--66 during
quiescence. The 0.1--2.4 keV luminosity is almost constant and in the
range $3-10\times 10^{34}$\ergs. Previous upper limits obtained with
Einstein IPC were of $\sim 5\times10^{35}$\ergs\ in the range 0.15--4.5
keV (Long et al. 1981). On Oct 7, 1993 the PSPC flux increased by
about an order of magnitude. This event took place at the same
orbital phase as the one reported by Corbet et al. (1995) and it is
likely to occur at the periastron passage (Skinner 1981).
These events are also similar to the weak outbursts reported by
Mavromatakis \& Haberl (1993). 

The quiescent X--ray luminosity can hardly derive from accretion onto the
neutron star surface because the required magnetic field is likely too low
(see above). A better interpretation is that matter is stopped at
the magnetospheric boundary by the centrifugal barrier, releasing
gravitational energy up to $r_{\rm m}$. Therefore, A0538--66 is likely 
the first object detected in this propeller regime. 

For this source the X--ray spectrum is also available. Thermal models
provide almost the same temperature during the weak
outburst (both the one reported in this paper and the two discovered
during the ROSAT all-sky survey) and in quiescence with $T_{\rm bb} \sim
0.2$ keV. 

The black body radius can be determined only for the observation with the
highest count rate and, on that occasion, it is much larger than the
neutron star radius and consistent with the magnetospheric radius: 
$R_{\rm bb}\gsim 5\times 10^6$ cm.
Thermal emission from a standard disk ``truncated" at the magnetospheric
boundary may explain the softness of the spectrum and the large black
body radius observed in A0538--66 during quiescence. 

The accumulation of material outside the magnetospheric boundary has also
been proposed to explain the soft thermal spectrum observed with the MPC
onboard Einstein, during strong outburst (Ponman et al. 1984). Thermal
emission from such ``shells" has also been invoked for Her X-1 (McCray \&
Lamb 1976) and SMC X-1 (Bunner \& Sanders 1979). During strong outbursts
matter can fall down to the neutron star surface heating this shell
($T_{\rm bb}\sim 2$ keV); on the contrary during weak outbursts or in
quiescence matter is likely stopped at $r_{\rm m}$. The lack of the
underlying hard X--ray spectrum results in a much smaller temperature
$T_{\rm bb}\sim 0.2$ keV, as testified by the lack of significant
differences in the PSPC spectra during quiescence or weak outbursts.

Observations with the {\it ASCA} and {\it SAX} satellites could better
constrain the emission properties of A0538--66 during quiescence. 

\begin{acknowledgements} 
I thank D. Lazzati, G.L. Israel, L. Stella and S. 
Mereghetti for helpful discussions. This work was partially supported by
ASI.
\end{acknowledgements}

\end{document}